\documentclass[10pt,a4paper,twocolumn]{article}

\pdfoutput=1

\usepackage{cite}

\usepackage{amssymb}

\usepackage{graphicx}

\usepackage{array}

\usepackage[font=footnotesize]{subfig}

\usepackage{paralist}

\usepackage[breaklinks=true]{hyperref}

\begin{document}

\title{State Machine Flattening: Mapping Study and Assessment}

\author{
Xavier~Devroey, Gilles~Perrouin\footnote{FNRS Postdoctoral Researcher}, Maxime Cordy\footnote{FNRS Research Fellow}, \\
Pierre-Yves Schobbens, Patrick Heymans\\
PReCISE, UNamur, Belgium\\~\\
Axel Legay \\
INRIA Rennes, Bretagne Atlantique, France
}

\maketitle

\begin{abstract}
State machine formalisms equipped with hierarchy and parallelism allow to compactly model complex system behaviours. Such models can then be transformed into executable code or inputs for model-based testing and verification techniques. Generated artifacts are mostly flat descriptions of system behaviour. \emph{Flattening} is thus an essential step of these transformations.  To assess the importance of flattening, we have defined and applied a systematic mapping process and 30 publications were finally selected.  However, it appeared that flattening is rarely the sole focus of the publications and that care devoted to the description and validation of flattening techniques varies greatly. Preliminary assessment of associated tool support indicated limited tool availability and scalability on challenging models. We see this initial investigation as a first step towards generic flattening techniques and scalable tool support, cornerstones of reliable model-based behavioural development.
~
\end{abstract}

\section{Introduction}

\emph{State machines} are popular models of system behaviour.  By providing them with a formal semantics, one can perform automated behavioural analysis (e.g. by model  checking
or model-based testing
) and code generation. 
In order to model complex systems in a concise and comprehensible manner, state machines have been equipped with various abstraction constructs such as hierarchy and parallelism~\cite{Harel1987}. Yet, abstraction comes with the cost of more elaborated semantics and potential ambiguities (e.g. in UML), thus preventing the direct use of automated analysis and generation tools.    

\emph{Flattening}~\cite{Harel1987}  -- a procedure that systematically transforms hierarchical state machines into state machines where all states are atomic -- was proposed as an answer. It bridges succinct modelling with formal semantics and automated analysis, allowing to envision end-to-end model-driven validation chains for complex systems~\cite{Devroey2012}. Flattening plays a pivotal role in behavioural analysis of software systems. Hence, its role in model-basel development and validation should be fully understood.             

In spite of its importance and widespread use, there has been no systematic effort to categorize flattening approaches and their applicability. This paper is a first step in this direction. We examine almost 20 years of scientific literature and perform a systematic mapping study~\cite{Petersen2008}. We follow the systematic approach used in the medical field~\cite{Kitchenham2007}, which is more appropriate for categorization purpose than systematic literature reviews \cite{Kitchenham2007,Petersen2008}. We nevertheless incorporated some relevant elements of systematic literature reviews as suggested by Petersen \textit{et al.}~\cite{Petersen2008}.
After an initial search that returned 167 publications, 30 of them were finally considered as relevant for the mapping. Our mapping relies on 4 dimensions (also called facets) covering research purpose, input/output models or the type of publication where the flattening techniques are applied or described. Our findings exhibit a balanced distribution of flattening use cases between validation and code generation purposes.  We also demonstrate that  flattening techniques are generally not described thoroughly, for these are often but a minor step of a larger process. Finally, the validation of the flattening technique, although essential to gain confidence in the engulfing approach, is insufficiently addressed. This latter point is supported by preliminary experimentation indicating that only a small number hierarchy and parallelism levels may be supported.         

The remainder of this paper is organised as follows. Section \ref{mapping-process} presents our mapping study process. Section \ref{results-discussion} presents the systematic map and discusses the results. Section \ref{experiment} describes our tool assessment and Section \ref{threats} covers threats to our empirical evaluation. Section \ref{conclusion} wraps up with conclusions and future research directions.  A companion webpage including all details of the mapping is available: \url{https://staff.info.unamur.be/xde/mappingstudy/}.

\section{Systematic Mapping Process} \label{mapping-process}


\begin{figure*}[t]
\centering
	\includegraphics[width=\textwidth]{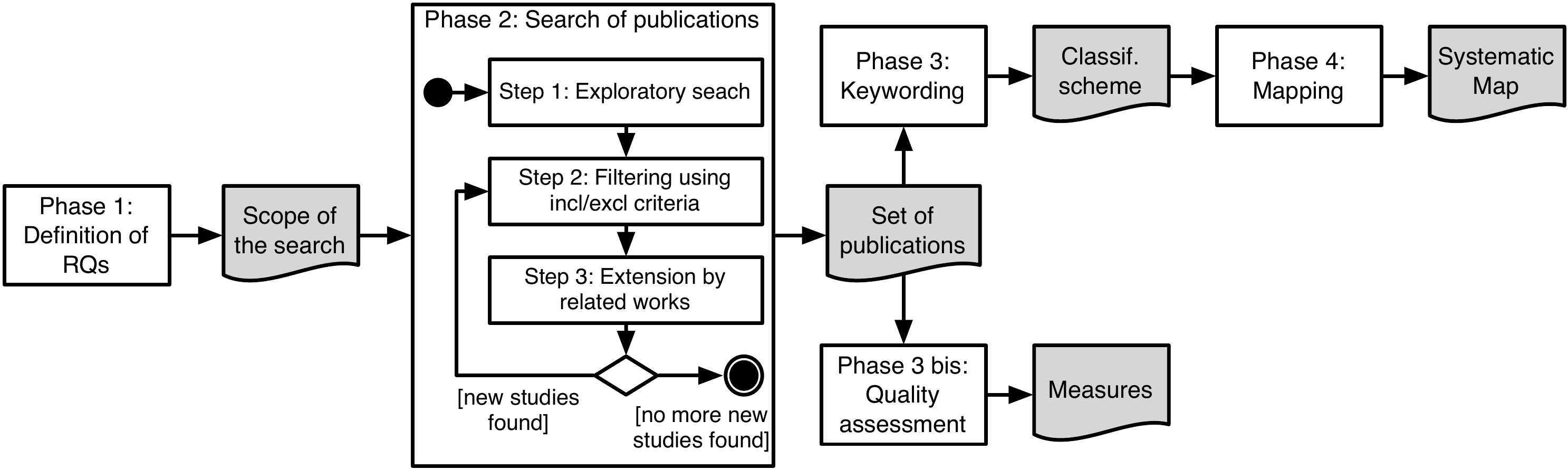}
	\caption{Systematic mapping process}
	\label{fig_systematic_mapping_process}
\end{figure*}

The definition of our systematic mapping process is inspired by \cite{DaMotaSilveiraNeto2011,engstrom2010software,Petersen2008}. However, as suggested by Petersen et al.~\cite{Petersen2008}, the process presented in Figure \ref{fig_systematic_mapping_process} and detailed hereafter does not strictly follow the classical systematic mapping review process. It incorporates practices of systematic literature reviews methods: the depth of reading is not limited to the abstract of the publications but rather adapted according to the importance of flattening in the publication; a quality assessment phase (see Section \ref{results-discussion}) has been added to evaluate the quality of the flattening description in the publications. 

\textbf{Phase 1: Research questions. }
%
The first phase is the definition of the research questions. They help delimiting the scope of the considered publications and allow to derive the search strings for publications exploration in phase 2. There are two points covered in this mapping study: the first one examines flattening techniques suited to eliminate hierarchy and parallelism (orthogonality) from a state machine-like model while the second one looks at the application contexts of such transformations. To cover flattening techniques, 2 research questions with a practical perspective are defined: (\emph{RQ1}) \emph{What are the input and output models used in the different flattening approaches?} (\emph{RQ2}) \emph{Do the different approaches support hierarchy (composite states) and parallelism (orthogonal states) in the input model?}
Flattening application context is covered by our last research question: (\emph{RQ3}) \emph{In which context are the different flattening approaches performed (e.g., code generation, test-case generation, semantic definition, etc.)?}



\textbf{Phase 2: Search of publications. }
%
In the second phase of the process, we gather relevant publications used to build the systematic map. To that aim, we follow the strategy presented in Figure \ref{fig_systematic_mapping_process}. First, we explore electronic databases and gather a raw set of publications. Next, we filter this raw set, and consequently obtain an initial set of relevant publications. From this initial set, we search for related (i.e. cited) work. The discovered papers are filtered and then added to the set of considered publications. We repeat the process until no new publication is added.
%
%
To perform the exploratory search in the electronic databases, we defined search strings designed to answer our research questions and inspired from four publications known by the team and experts of the domain \cite{Gogolla1998,Holt2010,Kalnins2005,Wasowski2004}. The considered databases with the different search strings are:
\begin{enumerate}
\item Google Scholar (\url{http://scholar.google.be/}):\\ \texttt{(computer science)} \texttt{AND} \texttt{(intitle:(} \texttt{state} \texttt{AND} \texttt{(machine} \texttt{OR} \texttt{machines} \texttt{OR} \texttt{chart} \texttt{OR} \texttt{charts)))} \texttt{AND} \texttt{(flattening~)} \texttt{AND} \texttt{(orthogonal} \texttt{OR} \texttt{parallel)}
\item Science Direct (\url{http://www.sciencedirect.com}):\\ \texttt{pub-date} \texttt{>} \texttt{2004} \texttt{and} \texttt{(("state} \texttt{machine"} \texttt{OR} \texttt{"state} \texttt{charts")} \texttt{AND} \texttt{(flat)} \texttt{AND} \texttt{(orthogonal} \texttt{OR} \texttt{parallel))} \texttt{and} \texttt{TITLE(state)[All} \texttt{Sources(Computer Science)]}
\item Computer Science Bibliographies (\url{http://liinwww.ira.uka.de/bibliography/}):\\ \texttt{+(flat flattening) +state +(diagram? chart?)}
\end{enumerate}
The initial search string used for Google Scholar did not contain the \texttt{computer science} keywords which lead to a lot of irrelevant results, most of them related to chemistry. We initially restricted ourselves to the [2005-2012] period to gather approaches compatible with the current version of UML.\footnote{UML2 was released in 2005. Please note that only the \emph{initial} scope of the publications is limited to the period 2005-2012. Our search process iteratively expands that period and eventually allows to consider ``non UML'' papers, e.g. \cite{Gogolla1998,Kansomkeat2003,Kim1999,Kuske2001,Wasowski2004}.} We found 167 publications in Google Scholar, 9 in Science Direct and 39 in Computer Science Bibliographies. The four publications known by the team and domain experts were contained in the 167 results returned by Google Scholar. This tends to indicate the relevancy of our search strings.
In the second step, we filter the result according to the following inclusion and exclusion criteria:

\emph{Inclusion criteria.} Books, articles, proceedings, technical reports and grey literature presenting a flattening technique with a hierarchical and / or orthogonal state machine or similar input (e.g., Harel Statecharts \cite{Harel1987}, etc.) and a flat state machine or assimilate as output (e.g., Finite State Machine (FSM), etc.)  are included. We also consider the publications where the produced output is source code, as source code may be used for testing and verification as well.

\emph{Exclusion criteria.} Literature only available in the form of presentation slides, publications where a flattening technique is only mentioned without details and publications citing a flattening technique described in another paper is excluded. In this last case, the cited papers are nonetheless considered according to the same inclusion and exclusion rules.

Once filtered, our set was extended in the third step by including papers cited in the publications. If the publication is focused on flattening, the references are picked up by screening the introduction, the background and related work parts. If not, only the ``flattening related part'', found by performing a word search in the documents, is considered for the references search. The considered regular expressions for the word search were: $flat.*$; $hierarch.*$; $orthogonal.*$. We repeat Steps 2 and 3 until our set of publications does not change.

After an initial filtering, we obtained an initial set of 24 publications \cite{Auer2009,Badreddin2012,Binder1999,Brajnik2011,Briand2005,David2001,David2002,Gogolla1998,Holt2010,Ipate2008a,Kalnins2005,Kansomkeat2003,Kim1999,Kuske2001,Masiero1994,Posse2010,Sacha2007,Schattkowsky2005,Schwarzl2010,Wasowski2004,Wasowski2005,Weissleder2009,Yao2006,Zoubeyr2010}. A first execution of Step 3 gave us 28 new papers. Among those, only 12 met the inclusion and exclusion criteria \cite{Agrawal2003,Ali1999,Andrea1994,Baresi2001,Bjorklund2001,Bond2001,Diethers2004,Engels2000,Hong2000,Minas2008,Riebisch2003,Roubtsova2001}. A second application of Step 3 gave us only one new publication, which did not match the inclusion and exclusion criteria. We eventually obtained a total of 36 publications.

\textbf{Phase 3: Keywording and Mapping. }
%
\begin{table}
\centering
\caption{Research Focus Facet (RQ3)}
\label{tab_researchNatureFacetCategores}
\begin{tabular}{|p{0.10\textwidth}|p{0.33\textwidth}|}
\hline 
\textbf{Category} & \textbf{Description} \\ 
\hline 
Code generation                                       & A model-driven approach generate or annotate source code from a flat state machine.\\ 
\hline 
Model checking                                        & A model to check is first flattened and the result is used as input of the model checker. \\ 
\hline 
Formal semantics                                   & The semantics of a state machine language is given as a transformation of which flattening is a step.\\
\hline
Model-based testing                                  & Test-cases are generated from a flat state machine.\\ 
\hline 
Flattening                            & Flattening is studied outside the scope of a specific application. \\ 
\hline 
Example       &  Flattening illustrates the use of a particular transformation framework. \\ 
\hline 
\end{tabular}
\end{table}
The classification scheme follows a two-step keywording process inspired by that of Petersen \textit{et al.}~\cite{Petersen2008}. In the first step, each paper is screened and tagged with keywords representing its contribution. In the second step, the classification scheme is built by grouping and combining the different keywords in higher level categories. Contrary to~\cite{Petersen2008}, where the considered publications are focused on the subject of the mapping study, we also consider papers where hierarchical / orthogonal models flattening is not the main aspect of the publication but only a step in a more general process. To deal with such cases, we propose to: (1) read only sections (adaptive reading depth~\cite{Petersen2008}) of publications where the model flattening aspect is explained and (2) guide the keywording process by our research questions:
\begin{inparaenum}[a)]
\item The purpose of the research (\emph{RQ3}).
\item The input model of the transformation (\emph{RQ1}).
\item The output model of the transformation (\emph{RQ1}).
\item Does the transformation support hierarchy / orthogonality in the input model? (\emph{RQ2}).
\item The implementation of the transformation (\emph{RQ3}).
\end{inparaenum}
In order to reduce bias, the first step of the keywording process has been done in parallel by two reviewers. The reviewers associate keywords with each publication, compare their results and discuss the differences. If they cannot agree on a given paper, a third reviewer solved the conflict. In our case, this happened for two papers: \cite{Wasowski2005,Minas2008}.

\begin{table}
\centering
\caption{Input Model Facet (RQ1, RQ2)}
\label{tab_inutModelCategories}
\begin{tabular}{|p{0.10\textwidth}|p{0.33\textwidth}|}
\hline 
\textbf{Category} & \textbf{Description} \\ 
\hline 
UML state machine                &  State machines built according to (any version of) the UML standard. \\ 
\hline 
Hierarchical Finite State Machine                            &  Hierarchical models based on state machines (e.g., Harel statechart \cite{Harel1987}). \\ 
\hline 
Hierarchical Timed Automata                    & Hierarchical state machines enriched with time information.  \\ 
\hline 
\end{tabular}
\end{table}

Next, the classification scheme is built by clustering the keywords into different categories. Similar categories are grouped to form what is called a facet. This is an iterative process where the classification scheme is enriched with each newly considered publication. In our case, four facets compose the classification scheme. The first facet (see Table~\ref{tab_researchNatureFacetCategores}) is concerned with the focus of the research described in the publication. This characterizes the broader context in which the flattening transformation is used. 
The second and third facets (see Tables \ref{tab_inutModelCategories} and \ref{tab_outputModelCategories}, respectively) describe the formalisms of the input and output models (respectively) employed by the flattening techniques described in the different publications.
The last facet (see Table~\ref{tab_researchFacetCategories}) classifies publications according to their type. These types range from problem-oriented papers (opinion, philosophical paper) to solution-oriented papers at various stages of their maturity.    

\begin{table}
\centering
\caption{Output Model Facet (RQ1, RQ2)}
\label{tab_outputModelCategories}
\begin{tabular}{|p{0.10\textwidth}|p{0.33\textwidth}|}
\hline 
\textbf{Category} & \textbf{Description} \\ 
\hline 
Flat UML state machine                &  Flat state machines based on any version of UML.\\ 
\hline 
Source code                & Code issued from a programming language or a textual specifications with a formal executable semantics.\\ 
\hline 
Model checker specification        & Any model checker specification, \textit{e.g.}, UPPAAL automata \cite{Bengtsson:1995kx} or Mealy machines \cite{Mealy:1955fk}.   \\ 
\hline 
Finite State Machine (FSM)        & This facet regroups the publications where the flattening transformation produces a flat FSM which is not a UML state machine. For instance: EFSM, Harel statechart, Symbolic transition system.  \\ 
\hline 
Graph                      &  Any kind of graph that other than finite state machine, \textit{e.g.}, petri net or testing flow graph \cite{Kansomkeat2003}.\\ 
\hline 
\end{tabular}
\end{table}

Once the classification scheme is defined, all the publications are classified. Despite our inclusion / exclusion criteria, we still found publications which, after complete review, were irrelevant with regards to our research questions: Brajnik \cite{Brajnik2011} presents the flattening proposed by Wasowski \cite{Wasowski2004}; Briand \textit{et al.} \cite{Briand2005} briefly discuss flattening in the related work part; Masiero and Maldonado \cite{Masiero1994} present a way to produce a reachability tree for hierarchical state machines which can not be considered as a flattening; in \cite{Posse2010} Posse preserves the hierarchical aspect of state machine in its target model; Zoubeyr \textit{et al.} \cite{Zoubeyr2010} do not describe any flattening technique; Engels \textit{et al.} \cite{Engels2000} describe a flattening of UML class hierarchy. All those publications match the exclusion criteria (flattening technique is only mentioned or comes from another paper) but were not detected earlier because of seemingly ``too broad" regular expressions. Irrelevant papers have been removed. Our final selection consists of 30 classified publications.

\begin{table}
\centering
\caption{Research Type Facet (RQ3) \cite{Petersen2008}}
\label{tab_researchFacetCategories}
\begin{tabular}{|p{0.10\textwidth}|p{0.33\textwidth}|}
\hline 
Category & Description \\ 
\hline 
Validation Research     &     Techniques investigated are novel and have not yet been implemented in practice. Techniques used are for example experiments, i.e., work done in the lab. \\ 
\hline 
Evaluation Research     &     Techniques are implemented in practice and an evaluation of the techniques is conducted. That means, it is shown how the technique is implemented in practice (solution implementation) and what are the consequences of the implementation in terms of benefits and drawbacks (implementation evaluation). This also includes identifying problems in industry. \\ 
\hline 
Solution Proposal         &     A solution for a problem is proposed, the solution can be either novel or a significant extension of an existing technique. The potential benefits and the applicability of the solution is shown by a small example or a good line of argumentation. \\ 
\hline 
Philosophical Papers     &     These papers sketch a new way of looking at existing things by structuring the field in form of a taxonomy or conceptual framework. \\ 
\hline 
Opinion Papers             &     These papers express the personal opinion of somebody whether a certain technique is good or bad, or how things should been done. They do not rely on related work and research methodologies. \\ 
\hline 
Experience Papers         &     Experience papers explain on what and how something has been done in practice. It has to be the personal experience of the author. \\ 
\hline 
\end{tabular}
\end{table}

\textbf{Phase 3 bis: Quality Assessment. }
%
In parallel with Phase 3, we propose, as suggested by da Mota Silveira Neto \textit{et al.}~\cite{DaMotaSilveiraNeto2011} to assess the quality of the selected publications using two groups of quality criteria. 
Our evaluation does not focus on the quality of the transformations themselves but rather on the quality of its description in the publications.
The first group evaluates the usability of the flattening technique in the publication:
\begin{inparaenum}
    \item[G]1.1 Is there a tool implementation?
    \item[G]1.2 Is there a small example?
    \item[G]1.3 Is there a more significant case study (even if not fully detailed)?
    \item[G]1.4 Are the input and output models described?
    \item[G]1.5 Does the publication present the limitations of the transformation?
\end{inparaenum}

The second group evaluates the degree of generality of the flattening process in the publication:
\begin{inparaenum}
    \item[G]2.1 Are there guidelines for the transformation separated from the example of the transformation (if any)?
    \item[G]2.2 Does it detail the transformation process for all the constructs of the input model?
    \item[G]2.3 Does the flattening technique support hierarchy?
    \item[G]2.4 Does the flattening technique support orthogonality?
\end{inparaenum}
A seemingly more objective metric is the ratio of text lines dedicated to the flattening process. Yet, it is rather cumbersome to perform and may be irrelevant since (1) longer descriptions are not necessarily more precise and complete, and (2) the exact value of this ratio may vary upon the reviewers.

\begin{table*}
\centering
\caption{Quality assessment results}
\label{tab_QARes}
\begin{tabular}{| p{2.5cm} | *{5}{c} | *{4}{c} | l | }
\hline
\textbf{Study ref.*}  & \textbf{G1.1} & \textbf{G1.2} & \textbf{G1.3} & \textbf{G1.4} & \textbf{G1.5} & \textbf{G2.1} & \textbf{G2.2} & \textbf{G2.3} & \textbf{G2.4} & \textbf{Type} \\
\hline
\multicolumn{11}{|l|}{Initial set} \\
\hline
Auer~\cite{Auer2009}          & \checkmark   & \checkmark   &    & \checkmark   & \checkmark   & \checkmark   & \checkmark   & \checkmark   & \checkmark   & T. rep.  \\
Badreddin~\cite{Badreddin2012}  & \checkmark   & \checkmark   & \checkmark   & \checkmark   & \checkmark   & \checkmark   & \checkmark   & \checkmark   & \checkmark  & Thesis \\
Binder~\cite{Binder1999}        &    & \checkmark   &    & \checkmark   & \checkmark   & \checkmark   & \checkmark   & \checkmark   & \checkmark  & Book \\
David~\cite{David2001}         & \checkmark   & \checkmark   & \checkmark   & \checkmark   & \checkmark   & \checkmark   & \checkmark   & \checkmark   & \checkmark   & T. rep. \\
David~\cite{David2002}         & \checkmark   & \checkmark   & \checkmark   & \checkmark   & \checkmark   & \checkmark   & \checkmark   & \checkmark   & \checkmark  & Art. \\
Gogolla~\cite{Gogolla1998}      &    & \checkmark   &    & \checkmark   & \checkmark   & \checkmark   & \checkmark   & \checkmark   &   & Art. \\
Holt~\cite{Holt2010}           & \checkmark   & \checkmark   &    & \checkmark   & \checkmark   & \checkmark   & \checkmark   & \checkmark   & \checkmark  & T. rep. \\
Ipate~\cite{Ipate2008a}        &    & \checkmark   &    & \checkmark   &    &    &    & \checkmark   & \checkmark  & Art. \\
Kalnins~\cite{Kalnins2005}      & \checkmark   & \checkmark   &    & \checkmark   & \checkmark   & \checkmark   & \checkmark   & \checkmark   & \checkmark  & Art. \\
Kansomkeat~\cite{Kansomkeat2003}  &    & \checkmark   &    & \checkmark   & \checkmark   & \checkmark   & \checkmark   & \checkmark   &  &  Proc. \\
Kim~\cite{Kim1999}          &    & \checkmark   &    & \checkmark   & \checkmark   & \checkmark   & \checkmark   & \checkmark   & \checkmark  & Art. \\
Kuske~\cite{Kuske2001}       &    & \checkmark   &    & \checkmark   & \checkmark   & \checkmark   & \checkmark   & \checkmark   & \checkmark  & Art. \\
Sacha~\cite{Sacha2007}      &    & \checkmark   & \checkmark   & \checkmark   & \checkmark   & \checkmark   & \checkmark   & \checkmark   &    & Art. \\
Schattkowsky~\cite{Schattkowsky2005} &    & \checkmark   &    & \checkmark   &    & \checkmark   &    & \checkmark   & \checkmark &  Art. \\
Schwarzl~\cite{Schwarzl2010}  &    &    &    & \checkmark   &    &    &    & \checkmark   &   & Proc. \\
Wasowski~\cite{Wasowski2004}  & \checkmark   & \checkmark   & \checkmark   & \checkmark   & \checkmark   & \checkmark   & \checkmark   & \checkmark   & \checkmark  & Art. \\
Wasowski~\cite{Wasowski2005}  &    & \checkmark   &    & \checkmark   & \checkmark   & \checkmark   & \checkmark   & \checkmark   & \checkmark &  Art. \\
Wei\ss{}leder~\cite{Weissleder2009}  & \checkmark   & \checkmark   &    &    &    &    &    & \checkmark   &   & Proc. \\
Yao~\cite{Yao2006}                      & \checkmark   & \checkmark   & \checkmark   & \checkmark   & \checkmark   & \checkmark   & \checkmark   & \checkmark   & \checkmark  & Proc. \\
\hline
\multicolumn{11}{|l|}{Added after iteration 1} \\
\hline
Agrawal~\cite{Agrawal2003}      & \checkmark  & \checkmark  &  & \checkmark  &  & \checkmark  &  & \checkmark  & \checkmark   &  T. rep.  \\
Ali~\cite{Ali1999}                       &  & \checkmark  &  & \checkmark  & \checkmark  & \checkmark  & \checkmark  & \checkmark  & \checkmark   &  Proc.  \\
Andrea~\cite{Andrea1994}         &  & \checkmark  &  & \checkmark  & \checkmark  & \checkmark  & \checkmark  & \checkmark  & \checkmark   &  Proc.  \\
Baresi~\cite{Baresi2001}             &  & \checkmark  & \checkmark  & \checkmark  &   & \checkmark  &  &  &   &  B. sect.  \\
Bjorklund~\cite{Bjorklund2001}  & \checkmark  & \checkmark  &  & \checkmark  &  &  &  & \checkmark  & \checkmark   &  Proc.  \\
Bond~\cite{Bond2001}                & \checkmark  & \checkmark  &  &  &  &  &  & \checkmark  &   &  Proc.  \\
Diethers~\cite{Diethers2004}      & \checkmark  &  &  &  &  &  &  & \checkmark  &   &  B. sect.  \\
Hong~\cite{Hong2000}                & \checkmark  & \checkmark  &  & \checkmark  & \checkmark  & \checkmark  &  & \checkmark  & \checkmark   &  Art.  \\
Minas~\cite{Minas2008}               &  & \checkmark  & \checkmark  & \checkmark  & \checkmark  & \checkmark  & \checkmark  & \checkmark  & \checkmark   &  Art.  \\
Riebisch~\cite{Riebisch2003}       & \checkmark  &  &  & \checkmark  &  & \checkmark  &  & \checkmark  &   &  Proc.  \\
Roubtsova~\cite{Roubtsova2001} &  & \checkmark  &  & \checkmark  & \checkmark  & \checkmark  &  & \checkmark  & \checkmark   &  Proc.  \\
\hline
\textbf{Ratio of \checkmark}  & \textbf{50\%}  & \textbf{90\%}  &  \textbf{27\% } &  \textbf{90\%}  & \textbf{67\% } &  \textbf{80\%}  &  \textbf{60\% } &  \textbf{97\%}  & \textbf{70\%   } & \\
\hline 
\end{tabular}
\end{table*}

Results of quality assessment are presented in Table \ref{tab_QARes}. It turns out that the huge majority (all but \cite{Baresi2001}) of the publications agree to define ``flattening'' as the ``removal of hierarchy'' in a state machine. In 97\% of the cases (Q.2.3), the input model supports hierarchy of states, and 70\% also supports orthogonality in sub-states. The only publication present in the mapping without supporting hierarchical states is \cite{Baresi2001}. We believe this publication should have been discarded by the protocol. Baresi and Pezz{\'e} \cite{Baresi2001} define the semantics of state machines as high-level Petri-nets, but the notion of hierarchy is applied to classes (although not as explicitly as in \cite{Engels2000}) and not states. Yet, the method may be applicable on hierarchical state machines (see \cite{Ipate2008a}).

Only 60\% of the publications thoroughly explain which constructs of the input model are supported (Q.2.2). These publications include: three technical reports (T. rep.) \cite{Auer2009,David2001,Holt2010}, one PhD thesis \cite{Badreddin2012} and one book \cite{Binder1999}. Those kinds of publications typically allow for more space to provide details than articles (Art.), book sections (B. sect.) or proceedings (Proc.). Two of the three publications belonging to the ``Example of transformation framework'' facet, \cite{Kalnins2005,Minas2008}, are also included. The third publication \cite{Agrawal2003} does not explain the complete transformation in detail but rather focuses on its performance. In most other cases \cite{David2002,Gogolla1998,Kansomkeat2003,Kim1999,Kuske2001,Sacha2007,Yao2006,Ali1999,Andrea1994,Minas2008}, the input model is a simplified version of UML state machine (except for \cite{Andrea1994} which uses Harel statecharts as input) where most of the pseudo-states have not been taken into account. In all those cases, the limitations of the transformation are presented in the publication (Q.1.5: 67\%). The only flattening techniques which takes history pseudo-state into account is the one presented by Wasowski et al.~\cite{Wasowski2004,Wasowski2005}. Finally, the transformation proposed by David and M\"{o}ller \cite{David2001} uses Hierarchical Timed Automata as input. This formalism has a well-defined semantics, contrary to UML state machines.

 We observe that only a very low percentage of techniques are validated on real or consequent case studies (Q.1.3: 27\%). It is not surprising since most of the publications are solution proposals. Only half of the publications have a tool implementation (Q.1.1). 
90\% of the publications illustrate the transformation with examples (Q.1.2), 90\% describe the input and output models (Q.1.4), and 80\% provide more detailed guidelines. Only \cite{Schwarzl2010} and \cite{Diethers2004} give no example nor guideline. The former describes very briefly the transformation in terms of input and output models. The latter presents the specifications and an overview of a model-checking tool without discussing the flattening transformation or the input model.

\section{Phase 4: Mapping} \label{results-discussion}

\begin{figure}
  \centering
  \includegraphics[width=0.45\textwidth]{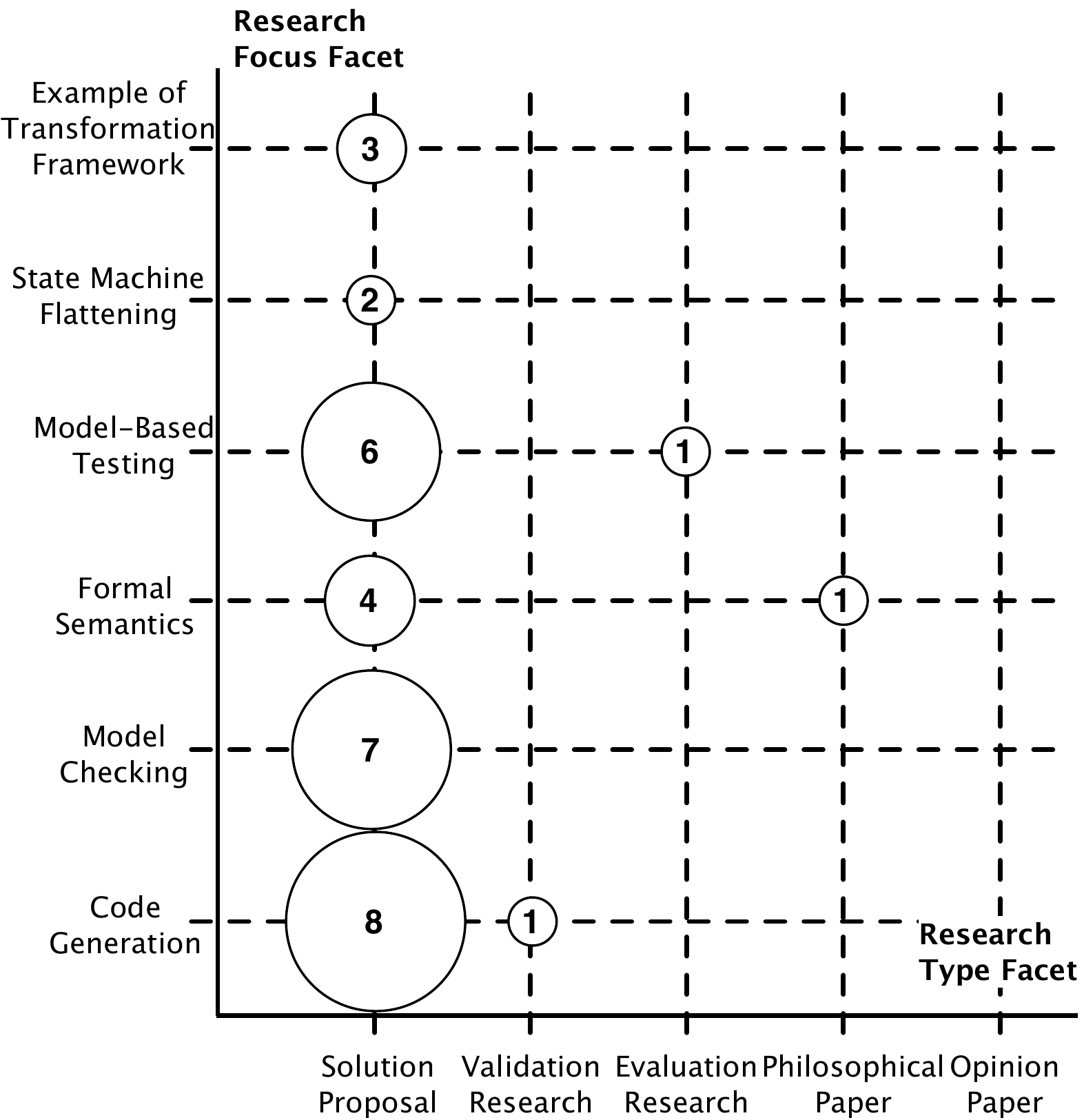}
  \caption{Systematic Map: Research type and focus facets}
  \label{fig_map_research_type_nature}
\end{figure}

\begin{figure}    
  \includegraphics[width=0.30\textwidth]{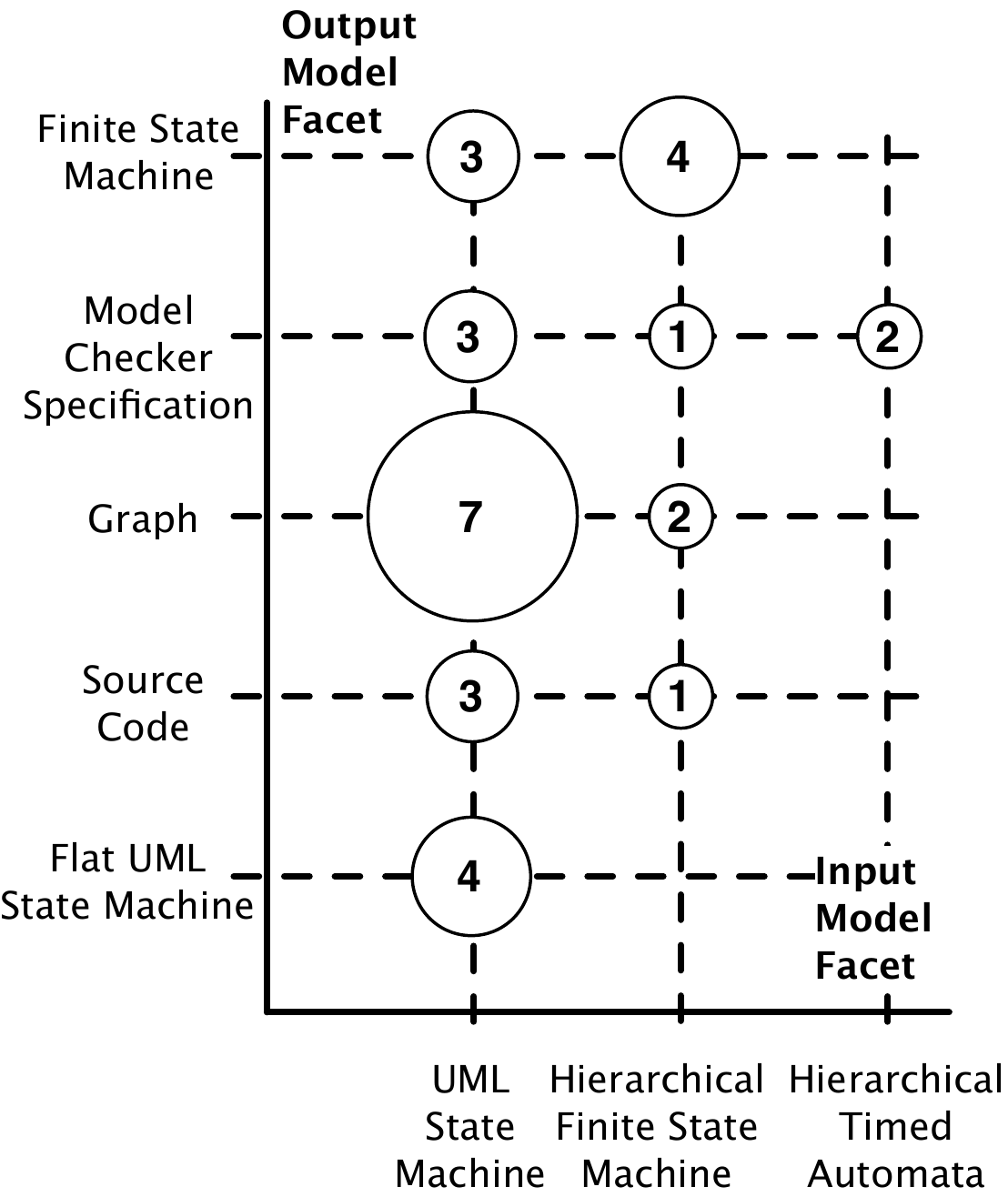}
  \caption{Systematic Map: Input and output facets}
  \label{fig_map_input_output}
\end{figure}

The complete mapping is not presented here due to space constraints. 
 Figures \ref{fig_map_research_type_nature} and \ref{fig_map_input_output} present a view of the mapping in the form of a bubble plot. The numbers in the bubbles represents the numbers of studies belonging to a particular combination of facets. In Figure \ref{fig_map_research_type_nature}, the number of studies is equal to 33. This is due to the classification of \cite{Badreddin2012}, \cite{David2001} and \cite{David2002}. \cite{Badreddin2012} is a PhD thesis classified as a validation research and a solution proposal in the research type facet. In \cite{David2001} and \cite{David2002} David \textit{et al.} uses flattening in order to generate code for a model checker, the reviewers agree to classify those two studies in both code generation and model checking in the research focus facets.


\textbf{Research Type (RQ3). }
Regarding the type of contributions where flattening considerations appear, the vast majority of them (93\%) are solution papers (see Figure \ref{fig_map_research_type_nature}). This is to be expected since flattening is a transformation used to bridge high-level models with existing lower level analysis tools and execution frameworks. We also note the poor level of validation and evaluation of the flattening algorithms (only 2 papers belong to the evaluation and validation research facets).  Among these two publications, only \cite{Weissleder2009} evaluated the effects of flattening on test case generation in practice. Finally, one paper \cite{Wasowski2005} considers a formal framework to discuss flattening algorithms complexity.   

\textbf{Research Focus (RQ3). }
The research focus facet illustrates a balanced distribution of the applications of flattening. The most common application of flattening is code generation (27\%). However this has to be mitigated by the fact that two publications \cite{David2002,David2001}  are producing code to be used with a model checker (UPPAAL).  Other kinds of generated code are dedicated to the synthesis of embedded systems \cite{Sacha2007,Bjorklund2001,Wasowski2004} or are instances of general purpose languages like JAVA \cite{Ali1999}. Model checking uses are related to consistency management \cite{Schwarzl2010,Yao2006,Diethers2004,Roubtsova2001} or IP telephony \cite{Bond2001}. Model checking exploits the fact that flattening is also a way to provide a formal semantics to hierarchical state machines \cite{Andrea1994,Gogolla1998,Kuske2001,Wasowski2005}. Three of these publications focus on UML state machines to make them analyzable. 
Model-based testing also extensively uses flattening approaches (23\% of the publications).  Unsurprisingly, most of the applications are centred on test case generation and selection \cite{Hong2000,Ipate2008a,Kansomkeat2003,Kim1999,Riebisch2003,Weissleder2009}. Rather than generating test cases over a flattened state machine, Ipate \cite{Ipate2008a} and Binder~\cite{Binder1999} propose to refine test cases gradually as states are decomposed. They argue that this incremental approach better copes with complexity. Riebisch \textit{et al.}~\cite{Riebisch2003} use state diagrams to refine UML use cases and subsequently generate tests at the system level. As for the other approaches, the generation algorithm requires a flat state machine. Two publications (\cite{Hong2000,Kim1999}) flatten Harel's statecharts and UML state machines (respectively) in order to generate flow graphs on which test-case generation and selection techniques are applied. Kansomkeat \textit{et al.}~\cite{Kansomkeat2003} flatten UML state machines to testing flow graphs in order to generate test-cases. Wei{\ss}leder \cite{Weissleder2009} flatten UML 2 state machines and uses coverage criteria to select and generate test-cases.
The two last categories (covering 17\% of the publications) concern the flattening transformation by itself. Holt \textit{et al.} \cite{Holt2010} describe in details a flattening algorithm implemented as a model transformation embedded in an Eclipse plug-in. As opposed to other publications \cite{Andrea1994,Schattkowsky2005,Kuske2001} based on graph transformations, the algorithm is given in an imperative manner. Finally, state machine flattening transformations are sometimes given as illustrative examples of model transformation frameworks \cite{Agrawal2003,Kalnins2005,Minas2008}.          

\textbf{Input and Output Facets (RQ1, RQ2). }
UML models are the most common input to flattening algorithms (67\%). There are, however, disparities in the supported UML constructs. The output of a flattening algorithm mainly depends on the goal for which the techniques is used. Graphs are preferred for providing formal semantics whereas verification-related work generally provides specifications for a model checker. If we match inputs with outputs, we infer that flattening is essentially an \emph{exogenous} transformation (\textit{i.e.} where the target language is different from the source language): UML state machines are both input and output in only 20\% of the publications.

\section{Preliminary Tool Assessement} \label{experiment}

\begin{table}[t]
\centering
\caption{Results: Average execution time}
\label{tab_ToolsExpRes}
\begin{tabular}{| l | l | l | l | }
\hline
\textbf{Depth}  & \textbf{UMPLE} & \textbf{SM2LIME} & \textbf{SCOPE} \\
\hline
0 & 0,306 sec.  & 0,038 sec.  &  $<$ 0,001 sec. \\
1 & 0,384 sec.  & 0,040 sec.  &  0,012 sec.\\
2 & 0,510 sec.  & 0,050 sec.  &  0,012 sec. \\
3 & Error  & 2,726 sec.  & Error\\
4 & Error  & $>$ 24 hrs  & Error \\
\hline
\end{tabular}
\end{table}

Our mapping study revealed a certain interest of the community for the flattening problem. A significant number of solutions have also been provided. Yet, tools are available in only 50\% of the publications and validation remains rare. Practical questions concerning existing tool support naturally arise. Thus, we decided to conduct an additional assessment of available tool support in the form of experiments. In particular we focused on one particular question: \emph{How do the proposed techniques scale to models of increasing complexity?}   

\textbf{Selection. }
Amongst the 15 publications that include an implementation \cite{Auer2009,Badreddin2012,David2001,David2002,Holt2010,Kalnins2005,Wasowski2004,Weissleder2009,Yao2006,Agrawal2003,Bjorklund2001,Bond2001,Diethers2004,Hong2000,Riebisch2003}  only 5 tools could be found on the Internet \cite{Auer2009,Badreddin2012,Wasowski2004,Weissleder2009,Agrawal2003}. We picked the three of them that have a command-line interface: SCOPE \cite{Wasowski2004}, UMPLE \cite{Badreddin2012} and SM2LIME \cite{Auer2009}. To evaluate their performance, we fed them with a state machine example that we successively extend with an increasing number of composite and orthogonal states. Since we are interested in reusable flattening techniques and we want our experiments to be reproducible, we considered only publicly available tools and did not contact the authors to obtain their implementation. Moreover, we are aware that our evaluation does not cover every existing tool as some are not presented in an elicited publication. 

\textbf{Experiment Design. }
Input models of varying complexity were automatically generated as follows. We started from a simple state machine $sm_{0}$ as base model with an initial state $i$, a simple state $s$ and a final state $f$ and two transitions: one from $i$ to $s$ and one from $s$ to $f$ triggered by an event with zero parameters. This machine with no composite state has a depth equal to $0$. To produce state machine $sm_{k}$ with a depth equal to $k$ ($k \in \{0, 1, 2, \ldots, 10\}$), we replaced the simple states in $sm_{k-1}$ by a composite state with two orthogonal regions containing each one a $sm_{0}$ state machine. We run each tool on the ten input models and measure their execution time using the Unix time (\texttt{/usr/bin/time}) command available on a Linux machine (kernel version: \#61-Ubuntu SMP Tue Feb 19 12:39:51 UTC 2013) with a Intel Core i3 (3.10GHz) processor and 4GB of memory. To minimize effects due to other running processes, we repeated each experiment five times.

\textbf{Results \& Discussion. }
Table \ref{tab_ToolsExpRes} presents the average execution time of each tool. 
None of the three selected tools could achieve more than 3 levels of depth: \emph{UMPLE} exits with a syntax error although $sm_{3}$ is generated using the same procedure as $sm_{2}$;  \emph{SCOPE} exits with a memory corruption error at  $sm_{3}$; \emph{SM2LIME} could process a $sm_{3}$ state machine but with an execution time jumping from 0,050 to 2,726 and has an (extrapolated) execution time greater than 24 hours for a $sm_{4}$ input model. Although the input models look simple, they become increasingly challenging due to an exponential blow-up in the numbers of parallel regions and interleaving transitions. Moreover, the structure of our models impedes the use of various optimization (\emph{e.g.} eliminating superfluous state/transitions using guard analysis \cite{Wasowski2004}) and thus yield a sharper growth in complexity.  Thus, these models are not intended to reflect any real system; they are meant to measure the scalability of the proposed tools. Additionally, they are agnostic of semantic variations of the different formalisms \cite{Taleghani2006,Crane2005}. This allows for fair comparisons between the tools.
   

State explosion did not allow for fine-grained trend analysis as models grow. However, our experiments confirm conclusions drawn on the mapping study regarding limited availability (overall only 33\% of the tools are freely available) and suggest that hierarchy and parallelism threaten scalability. This further motivates the need for new efficient techniques. 


\section{Threats to validity} \label{threats}


\textbf{Publication bias. } We cannot guarantee that all relevant publications were selected, especially since the state machine flattening is rarely the main focus of the publications but rather a way to achieve a more general purpose. We tried to mitigate this threat by adopting an approach were the set of publications is built iteratively by including cited papers. The publication dates of the papers added at each step of publications search  (Phase 2)  shows a good coverage for a period from 1994 to 2012: in the initial set the oldest publication (\cite{Masiero1994}) was published in 1994 and the most recent (\cite{Badreddin2012}) was published in 2012; the publications added in iteration 1 were less recent (from 1994 \cite{Andrea1994} to 2008 \cite{Minas2008}) and the publication found during iteration 2 (and excluded from the set of publications) was published in 1996 (\cite{Harel1996a}). Moreover, the selected papers originate from different research areas, thus indicating that our selection procedure covers a large scope of publication. Finally, the cited documents of rejected publications were still included in the set at step 3 of the ``search of publications'' phase.

\textbf{Research strings. } The search strings used for database mining may have many synonyms. Relevant publications may thus remain undetected. Still, the used strings allowed us to successfully detect the four initial papers known by the domain experts. As for publication bias, the distribution of the publication dates from 1994 to 2012 shows that the initial publication period is no major threat to completeness.

\textbf{Keywording. } As the considered publications are not all focused on flattening, the keywording process may be influenced by the reviewer. To avoid bias, the keywording process is performed in parallel by two different readers. Once the keywords have been associated with the publications, the readers compare their results and discuss the differences between associated keywords. If conflicts between the associated keywords remain, a third party acts as an arbitrator. 

\textbf{Quality assessment. } As for the keywording process, the point of view of the reviewer may influence the answers to the different questions. To overcome this as much as possible, only yes/no answers are allowed. Since the quality assessment was performed in parallel with the keywording phase, the two reviewers have assessed the quality of the different publications. Again, divergences were solved by a third party.

\textbf{Tool Selection. }  The rationale behind tool selection was to assess whether tools mentioned in the publications were publicly available and ready for practical use.  While our answer is negative to these questions, efficient tools may have been missed because of our focus on scientific publications. This threat can be mitigated by the fact that proper documentation is necessary to understand the tool's input model formalism and thus generate models for experimentation. However, conducting a wider assessment on a larger set of tools is part of our research agenda.

\textbf{Model Complexity. } We created and systematically extended challenging models.  Such an approach is relevant to compare tools on a fair basis. To the best of our knowledge, there exists no survey about the size and complexity of state machines designed in industry. It is thus possible that such a high level of complexity never occurs in practice.  

\section{Conclusion} \label{conclusion}

Due to their compactness and formal semantics, state machines are a powerful means of modelling, verifying and validating the behaviour of complex systems. However, abstraction mechanisms such as composite and parallel states impede the use of automated analysis and generation techniques, often requiring flat structures.
Flattening is called to play a crucial role in bridging abstract models with analysable and executable ones. Recognising the lack of overall cartographies of flattening approaches, this systematic mapping study is a first step in this direction. In particular, we outlined a balanced status were flattening is used equally for model-based validation (testing, verification) and code generation. Flattening also barely appears to be an object of interest in itself but rather a step towards a more general objective.  This has impacts on the quality of the description of flattening algorithms.  First, precise constructs supported by the flattening transformation are not always provided, making the applicability of a given technique to a specific context difficult to evaluate. Second, the validity of the flattening transformation is barely addressed, which is necessary to gain confidence in the quality of the bridge. Mapping study conclusions are supported by our preliminary assessment of flattening tools that exhibited reliability and scalability issues on small but challenging models.  

In the future, we would like to provide a complete (including syntax and semantics concerns) taxonomy of flattening approaches.  This will enable the design of generic flattening techniques and tools. We will also offer a sound evaluation framework to compare flattening techniques and thus will help understanding in which situation(s) a given flattening approach is the most appropriate.  These are mandatory steps towards reliable, end-to-end, model-based behavioural development.    




\bibliographystyle{plain}
\bibliography{biblio}

\end{document}